\documentclass[aps,prd,a4paper,showpacs,twocolumn]{revtex4-1}
\usepackage{amsmath,amssymb,bm}
\usepackage[utf8]{inputenc}
\usepackage[T1]{fontenc}
\usepackage{gensymb}
\usepackage{nicefrac} 
\usepackage{array}
\usepackage{siunitx}
\usepackage{epsfig}
\usepackage{graphicx}
\usepackage{slashed}
\usepackage{epsfig} \usepackage{color}
\usepackage{mathrsfs} \usepackage{url}
\usepackage{subfigure}

\usepackage{xspace} 
\usepackage[bookmarks, breaklinks, colorlinks,urlcolor=black, citecolor=red, 
linkcolor=blue]{hyperref}

\usepackage{appendix}
\usepackage{lipsum}
\usepackage{bigints}
\usepackage{tikz}
\usepackage[compat=1.1.0]{tikz-feynman}
\usepackage{braket}
\usepackage{mathrsfs}

\def\tev{\ensuremath{\mathrm{\,Te\kern -0.1em V}}}
\def\gev{\ensuremath{\mathrm{\,Ge\kern -0.1em V}}}
\def\mev{\ensuremath{\mathrm{\,Me\kern -0.1em V}}}

\def\dsp{\displaystyle}
\def\nn {\nonumber}

\allowdisplaybreaks

\begin{document}

\title{Testing $WW\gamma$ vertex in radiative muon decay}

\author{Anirban Karan}\email{kanirban@imsc.res.in}
\affiliation{The Institute of Mathematical Sciences, Taramani, Chennai 
	600113, India}
\affiliation{Homi Bhabha National Institute, BARC Training School Complex, 
Anushaktinagar, Mumbai 400094, India }

\author{Rusa Mandal}\email{Rusa.Mandal@ific.uv.es}
\affiliation{ IFIC, Universitat de Val$\grave{e}$ncia-CSIC, Apt. Correus 22085, 
E-46071 Val$\grave{e}$ncia, Spain}

\author{Rahul Sinha}\email{sinha@imsc.res.in}
\affiliation{The Institute of Mathematical Sciences, Taramani, Chennai 
600113, India}
\affiliation{Homi Bhabha National Institute, BARC Training School Complex, 
Anushaktinagar, Mumbai 400094, India }

%
%
%

\date{\today}

\begin{abstract}
Large numbers of muons  will be produced at facilities developed to probe lepton
flavor violating process $\mu\to e\gamma$. We show that by constructing a
suitable asymmetry, radiative muon decay $\mu\to e \gamma\nu_\mu\bar{\nu}_e$ can
also be used to test the $WW\gamma$ vertex at such facilities. The process has
two missing neutrinos in the final state and on integrating their momenta, the
partial differential decay rate shows no radiation-amplitude-zero. We establish,
however, that an easily separable part of the normalized differential decay
rate, odd under the exchange of photon and electron energies, does have a zero
in the case of standard model (SM). This \emph{new type of zero} has hitherto
not been studied in literature. A suitably constructed asymmetry using this
fact, enables a sensitive probe for the $WW\gamma$ vertex beyond the SM. With a
simplistic analysis, we find that the $C$ and $P$ conserving dimension four
$WW\gamma$ vertex can be probed at ${\cal O}(10^{-2})$ with satisfactory
significance level.
\end{abstract} 

\maketitle 

\section{Introduction} 

The $SU(2)_L\otimes U(1)_Y$ theory of electroweak interactions has been tested
extensively in last few decades and there is no doubt that it is the correct  theory
at least up to a \tev scale. This conviction is largely based on the precision
measurements at LEP and the consistency of top and Higgs boson masses which
could be predicted taking radiative corrections into account. The gauge boson
and Higgs boson self interactions are, however, not as well probed either by
direct measurement or by radiative corrections and it is possible that some
deviations from the standard Model (SM) loop level values might still be seen.
To ascertain the validity of SM it is critical that the  $WW\gamma$ vertex, which is predicted uniquely in SM, be probed to an accuracy consistent with loop
level corrections to it. Several experiments~\cite{cms,atlas,
d0,cdf,delphi,aleph,opal,l3} have measured parameters that probe the $WW\gamma$
and $WWZ$ vertex, but the accuracy achieved is still insufficient to probe one
loop corrections to it within the SM.

In this paper, we have investigated how the  $C$ and $P$ conserving dimension
four $WW\gamma$ operator can be probed experimentally using radiative muon
decays.  The vertex factor for this operator is usually denoted by
$\kappa_\gamma$ and is uniquely predicted in the SM. At tree level
$\kappa_\gamma=1$ in the SM and the absolute value of the one loop corrections
to the tree level values of $\kappa_\gamma$ is restricted to be less than
$1.5\times10^{-2}$~\cite{courture}. However, the current global average
$\kappa_\gamma= 0.982\pm0.042$~\cite{Patrignani:2016xqp} has too large an
uncertainty to probe the SM up to one loop accuracy. Of the experimentally
measured values of $\kappa_\gamma$, only  ATLAS and CMS collaborations use the
data for real on-shell photon emission in hadron colliders \cite{cms,atlas},
probing the true magnetic moment of the $W$-boson.

One can expect $\kappa_\gamma$ to deviate from its SM value by only a few
percent, hence, we must choose the mode to be studied very carefully. Radiative
muon decay $\mu\to e \gamma\nu_\mu\bar{\nu}_e$ is a promising mode to measure
the true magnetic moment (due to real photon in the final state)  of the
$W$-boson in this regard. At first sight the measurement of $W$-boson gauge
coupling using low energy decay process may seem impossible, since the effect is
suppressed by two powers of the $W$-boson mass. The process has two missing
neutrinos in the final state and on integrating their momenta the partial
differential decay rate shows no radiation-amplitude
zero~\cite{Mikaelian:1979nr}. Moreover, the differential decay rate does not
show enough sensitivity to a deviation of the $WW\gamma$ vertex from that of the
SM. We show, however, that an easily separable part the normalized differential
decay rate (odd under the exchange of photon and electron energies) does have a
zero in the case of SM.  The vanishing of the odd contribution under the
exchange of final state electron and photon energies in the decay  rate is a
\emph{new type of zero},  hitherto not been studied in literature. A suitably
constructed asymmetry using this fact enables adequate sensitivity to probe the
$WW\gamma$ vertex beyond the SM. We consider a very restricted part of the phase
space where the asymmetry is larger than statistical errors for our study. Large
number of muons are expected to be produced  for COMET~\cite{COMET},
MEG~\cite{MEG} and Mu2e~\cite{Mu2e} collaborations to probe lepton flavor
violating processes like $\mu\to e\gamma$. The radiative muon decay  $\mu\to e
\gamma\nu_\mu\bar{\nu}_e$ \cite{kuno} discussed in this paper is the dominant background
process for this case. The large sample of $\mu\to e \gamma\nu_\mu\bar{\nu}_e$
produced at such facilities make them an ideal environment to probe $WW\gamma$
vertex, with reduced statistical uncertainty, as discussed in this paper. In a 
simulation using $\eta_\gamma\equiv\kappa_\gamma-1=0.01$, we find
that the asymmetry constructed by us, can probe this $\eta_\gamma$ value with a
$3.9\sigma$ significance.

The rest of the paper is organized as follows. In Sec.~\ref{Sec:theory} we
briefly discuss the decay kinematics and relevant expressions for decay rate.
These results are used to construct the observables in Sec.~\ref{Sec:obs}, where
we also explain why a zero in odd amplitude is expected.
Section.~\ref{Sec:simulation} deals with the numerical analysis to probe the
$WW\gamma$ vertex and finally we conclude in Sec.~\ref{Sec:conclusion}.

\section{Theoretical Framework}
\label{Sec:theory}

\begin{figure}[h]
\includegraphics[scale=0.25]{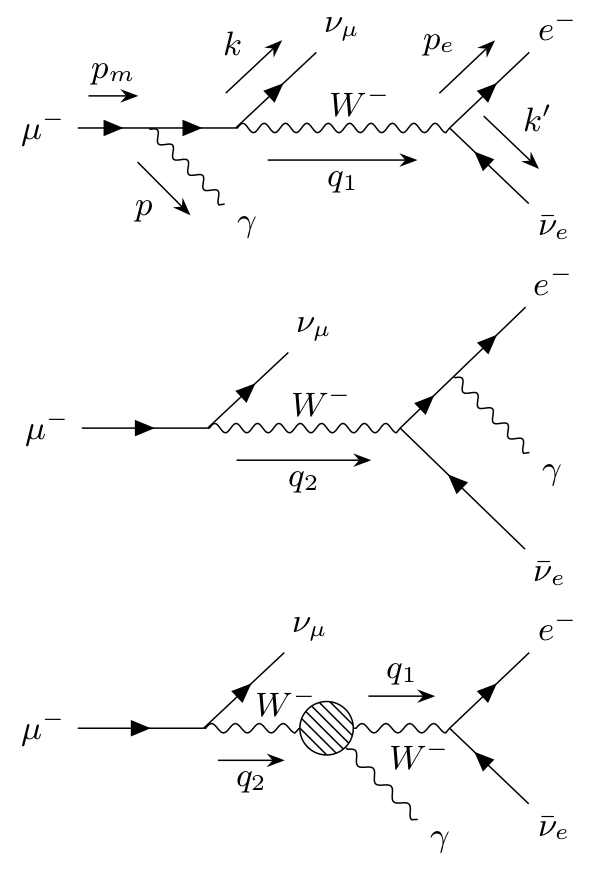}
\caption{Feynman diagrams for radiative muon decay.}
\label{feyn}
\end{figure}

\begin{figure}[h]
\begin{center}
\includegraphics[scale=0.25]{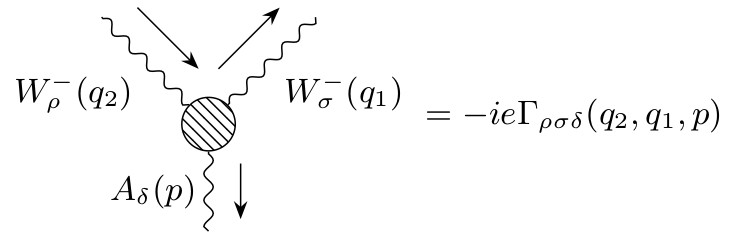}
\caption{Feynman rule for effective $WW\gamma$ vertex.}
\label{tbf}
\end{center}
\end{figure}

In this section we briefly discuss the theoretical set up for the radiative 
muon decay.
The radiative muon decay proceeds through three Feynman diagrams, shown in 
Fig.~\ref{feyn}, where the photon in the final state can either arise from any 
of the initial and final state leptons or the $W$ boson in the propagator. The 
later process is of our particular interest.
We define the four momenta of incoming $\mu^-$, outgoing
$e^-,\,\gamma,\,\nu_\mu,\,\bar{\nu}_e$ as  $p_m,\,p_e,\,p,\,k$ and $k'$,
respectively, and the masses of muon, electron and $W$-boson are denoted by
$m_\mu$, $m_e$ and $m_W$, respectively. The amplitudes corresponding to these
three diagrams (from top to bottom), labelled with subscript 1 to 3, can be
expressed as
\begin{align}
\label{M1}
i\mathcal{M}_1&=\big(\frac{-ieg^2}{8}\big)
\overline{u}(p_e)\gamma_\beta^{}(1-\gamma_5^{})v(k') 
\Bigg[\frac{g^{\alpha\beta}-\dsp\frac{q_1^\alpha 
q_1^\beta}{m_W^2}}{q_1^2-m_W^2}\Bigg]\nn \\  & \times 
\overline{u}(k)\gamma_\alpha^{}(1-\gamma_5^{})
\Big[\frac{1}{\slashed{p}_m-\slashed{p}-m_\mu}\Big]\gamma_\delta^{}
 u(p_m)\epsilon^{*\delta},\\
%
\label{M2}
i\mathcal{M}_2&=\big(\frac{-ieg^2}{8}\big)
\overline{u}(k)\gamma_\alpha^{}(1-\gamma_5^{})u(p_m) \Bigg[\frac{g^{\alpha\beta}-\dsp\frac{q_2^\alpha q_2^\beta}{m_W^2}}{q_2^2-m_W^2}\Bigg] \nn \\
& \times \overline{u}(p_e)\gamma_\delta^{} 
\Big[\frac{1}{\slashed{p}_e+\slashed{p}-m_e}\Big] 
\gamma_\beta^{}(1-\gamma_5^{}) v(k')\epsilon^{*\delta},\\
%
\label{M3}
i\mathcal{M}_3&=\big(\frac{-ieg^2}{8}\big)
\overline{u}(k)\gamma_\alpha^{}(1-\gamma_5^{})u(p_m) \Bigg[\frac{g^{\alpha\rho}-\dsp\frac{q_2^\alpha q_2^\rho}{m_W^2}}{q_2^2-m_W^2}\Bigg]\nn \\&
\times \Bigg[\frac{g^{\sigma\beta}-\dsp\frac{q_1^\sigma 
q_1^\beta}{m_W^2}}{q_1^2-m_W^2}\Bigg]\overline{u}(p_e)\gamma_\beta
(1-\gamma_5^{})v(k')\nn \\&
\times\Gamma_{\rho\sigma\delta}(q_2,q_1,p)\epsilon^{*\delta},
\end{align}
where 
$e$ and  $g$ are the charge of positron and weak coupling constant, respectively; $q_1^\mu=p_e^\mu+k'^\mu$ and $q_2^\mu=p_m^\mu-k^\mu$. In Eq.~\eqref{M3}, $\Gamma_{\rho\sigma\delta}(q_2,q_1,p)$ denotes the effective triple gauge boson vertex for electroweak interaction as shown in Fig. \ref{tbf} .

The most general couplings of $W$ to the neutral gauge bosons $\gamma$ and $Z$ can be described by the following effective Lagrangian \cite{hagiwara},
\begin{equation}
\label{eq:Lag}
\begin{split}
\mathcal{L}_{eff}^V=&-ig_V^{}[g_1^V(W^\dagger_{\mu\nu} W^\mu-W^{\dagger\mu} W_{\mu\nu})V^\nu\\&+\kappa_V^{}W^\dagger_\mu W_\nu V^{\mu\nu}+\frac{\lambda_V^{}}{m_W^2}W^\dagger_{\lambda\mu} W^\mu_\nu V^{\nu\lambda}\\&+if_4^V W^\dagger_\mu W_\nu (\partial^\mu V^\nu+\partial^\nu V^\mu)\\&-if_5^V\epsilon^{\mu\nu\rho\sigma}(W^\dagger_\mu \overset\leftrightarrow{\partial_\rho} W_\nu)V_\sigma\\&+\widetilde{\kappa}_V^{}W^\dagger_\mu W_\nu \widetilde{V}^{\mu\nu}+\frac{\widetilde{\lambda}_V^{}}{m_W^2}W^\dagger_{\lambda\mu} W^\mu_\nu \widetilde{V}^{\nu\lambda}].
\end{split}
\end{equation}
Here, $V$ corresponds to $\gamma$ or $Z$,  $g_\gamma^{}=e$ and 
$g_Z^{}=e\cot\theta_W$ where $\theta_W$ is the Weinberg angle.  
$W_{\mu\nu}=\partial_\mu 
W_\nu-\partial_\nu W_\mu$, $V_{\mu\nu}=\partial_\mu V_\nu-\partial_\nu V_\mu$, 
$\widetilde{V}_{\mu\nu}=\frac{1}{2}\epsilon_{\mu\nu\rho\sigma}V^{\rho\sigma}$, 
$(A\overset\leftrightarrow{\partial}_\mu B)=A(\partial_\mu B)-(\partial_\mu 
A)B$ and Bjorken-Drell metric is taken as 
$\epsilon_{0123}=-\epsilon^{0123}=+1$. In the SM, at tree level,
$g_1^V=\kappa_V=1$ and all other coupling parameters are zero. 

In the case of radiative muon decay, the vertex with $W$ boson pair and a photon
field is involved where among the seven coupling parameters, $f_4^\gamma$,
$\widetilde{\kappa}_\gamma^{}$ and  $\widetilde{\lambda}_\gamma^{}$ denote the
coupling strengths of $CP$ violating interactions in the Lagrangian (in
Eq.~\eqref{eq:Lag}) and are constrained to be less than $\sim
(10^{-4})$\cite{marciano} due to the measurements of neutron electric dipole
moment in case of direct $CP$ violation. 
Due to the $CP$ violating 
nature of these couplings, deviations from the SM  contributions are
proportional to square of these couplings and thus are highly suppressed, as compared to $CP$-conserving contributions. Hence, we neglect 
the $CP$ violating parameters for the rest of the discussion of the
paper. The demand of $C$ and $P$ to be conserved separately in the Lagrangian
allows us to choose vanishing $f_5^\gamma$. It is obvious that the muon
radiative decay will not be sensitive to the dimension six-operator involving
$\lambda_\gamma$, due to an additional $m_W^2$ suppression. The measurement of 
$\lambda_\gamma$ is possible only at high energy colliders. Hence, we can safely
neglect the deviation of $\lambda_\gamma$ from its SM value of zero.
Furthermore, the value of the coupling $g_1^\gamma$ is fixed to be unity due to
electromagnetic gauge invariance. Thus, in momentum space the $WW\gamma$ vertex
can be expressed as
\begin{equation}
\label{eq:vertex}
\begin{split}
\Gamma_{\rho\sigma\delta}&(q_2,q_1,p)=g_{\rho\sigma}(q_2+q_1)_\delta
+g_{\sigma\delta}(p-q_1)_\rho\\&-g_{\delta\rho}(p+q_2)_\sigma
+\eta_\gamma(p_\rho g_{\sigma\delta}-p_\sigma g_{\rho\delta}),
\end{split}
\end{equation}
where $\eta_\gamma\equiv\kappa_\gamma-1$ and $q_2,q_1,p$ are the four momenta 
of incoming $W^-$, outgoing $W^-$ and outgoing photon respectively, as depicted 
in Fig.~\ref{tbf}. 

It is apparent from Fig.~\ref{feyn} and Eqs.~\eqref{M1}-\eqref{M3}, that
amplitude ($\mathcal{M}_3$) containing effective vertex 
$\Gamma_{\rho\sigma\delta}$ is
$1/m_W^2$ suppressed compared to the other two contributions  $\mathcal{M}_1$
and $\mathcal{M}_2$. Hence, within the SM, the first two Feynman-diagrams in
Fig.~\ref{feyn} are sufficient to study the process. On the other hand only the 
third diagram is sensitive to
$\eta_\gamma$. Thus, in order to retain sensitivity to  $\eta_\gamma$ in
$\Gamma_{\rho\sigma\delta}$, it is necessary and sufficient to keep 
contributions up
to $\mathcal{O}(1/m_W^4)$, in the amplitudes. To achieve this we expand the $W$
boson propagator in the power series of $(q_j^2/m_W^2)$ as
\begin{equation}
\hspace*{-0.15cm}-i\Bigg[\frac{g^{\alpha\beta}-\dsp\frac{q_j^\alpha 
q_j^\beta}{m_W^2}}{q_j^2-m_W^2}\Bigg]\approx\frac{i}{m_W^2}
\Bigg[g^{\alpha\beta}+\frac{q_j^2}{m_W^2}\Big(g^{\alpha\beta}-\frac{q_j^\alpha
 q_j^\beta}{q_j^2}\Big)\Bigg].
\end{equation}
The total amplitude can be expressed as 
$\mathcal{M}=\mathcal{M}_1+\mathcal{M}_2+\mathcal{M}_3$
and we calculate differential cross section keeping  all the amplitudes up to
$\mathcal{O}(1/m_W^4)$. Since the neutrinos $\nu_\mu$ and $\bar{\nu}_e$ cannot
be observed we integrate the $\nu_\mu$ and $\bar{\nu}_e$ momenta, and define the
$\nu_\mu\bar{\nu}_e$ invariant momentum as $q$. As the decay now looks like a
3-body decay it is meaningful to define effective Mandelstam like variable
constructed from the invariant momentum square of $e^-\nu_\mu\bar{\nu}_e$ 
system as $t$ and
that of $\gamma\nu_\mu\bar{\nu}_e$ system as $u$. Hence, $(p_e+q)^2=t$ and
$(p_\gamma+q)^2=u$. Notice that, $q^2$ is not a constant for our decay.
 It is, however, much more convenient to 
define normalized parameters
\begin{equation}
\label{param}
\begin{split}
x_p=\frac{t+u}{2(q^2+m_\mu^2)},\\
y_p=\frac{t-u}{2(q^2+m_\mu^2)},\\
q_p^2=\frac{q^2}{(q^2+m_\mu^2)},
\end{split}
\end{equation}
which can be written in terms of the observable quantities, the photon energy 
$E_\gamma$, the electron energy $E_e$ and the angle between
the electron and photon $\theta$ as follows.
\begin{eqnarray}
& x_p=\dsp\frac{m_\mu(m_\mu-E_e-E_\gamma)}{2[m_\mu^2-E_\gamma m_\mu-E_e 
m_\mu+E_e E_\gamma (1-\cos\theta)]},\\[1.5ex]
& y_p=\dsp\frac{m_\mu(E_e-E_\gamma)}{2[m_\mu^2-E_\gamma m_\mu-E_e m_\mu+E_e 
E_\gamma (1-\cos\theta)]},\\[1.5ex]
&
\!q_p^2=\!\dsp\frac{m_\mu^2-2E_\gamma m_\mu-2E_e m_\mu+2E_e E_\gamma 
(1-\cos\theta)}{2[m_\mu^2-E_\gamma m_\mu-E_e m_\mu+E_e E_\gamma 
(1-\cos\theta)]}.
\end{eqnarray}
The parameters of interest for 
the derivation, $x_p$, $y_p$ and  $q_p^2$ can easily be inverted in terms of the
observables $E_e$, $E_\gamma$ and $\cos\theta$ as,  
\begin{align}
\label{Ee}
&E_e=\frac{m_\mu}{2}\bigg(\frac{1-q_p^2-x_p+y_p}{1-q_p^2}\bigg),\\
\label{Egamma}
&E_\gamma=\frac{m_\mu}{2}\bigg(\frac{1-q_p^2-x_p-y_p}{1-q_p^2}\bigg),\\
\label{costheta}
&\cos\theta=\frac{(q_p^2-x_p)^2+2x_p-y_p^2-1}{(1-q_p^2-x_p)^2-y_p^2}.
\end{align}
We notice that replacing $y_p$ by $-y_p$ while keeping $q_p^2$ and $x_p$ 
unchanged actually results in swapping the energies of photon and electron 
keeping the angle between them unaltered. This feature will play a very crucial 
role in defining the observable asymmetry in Sec.~\ref{Sec:obs}.

We have ignored the electron mass, $m_e$, starting from Eq.~\eqref{param}
as it results in significant simplification of analytic expressions. It is of
course well-known that neglecting the electron mass results in the persistence
of wrong helicity right-handed electron~\cite{Sehgal:2003mu,Schulz:2004xd} in
this decay as a result of inner bremsstrahlung from the electron (see second
diagram of Fig.~\ref{feyn}). The results are in obvious disagreement depending
on whether $m_e$ is retained or not. We will therefore very carefully consider
the issue of electron mass to justify the neglect of $m_e$ for our limited
purpose of extracting $\eta_\gamma$, while acknowledging that $m_e$ should not
be ignored in general. In order to retain maximum sensitivity to $\eta_\gamma$
the kinematic domain is chosen to minimize the soft photon and collinear
singularity contributions; the effect of $m_e$  is found to be insignificant in 
the kinematic domain sensitive to $\eta_\gamma$. Our
calculations have been verified retaining $m_e$ throughout. Critical expressions
including $m_e$ contributions are presented in Appendix~\ref{Sec:Appendix} for
clarity. Expressions for $x_p$ and $y_p$ are modified to accommodate effects of
$m_e$, while retaining an {\em apparent exchange symmetry} between $E_\gamma$
and $E_e$ under the newly defined variables $x_n$ and $y_n$ in 
Eq.~\eqref{eq:newxy}.

We consider only the normalized differential decay rate 
$\overline{\Gamma}(x_p,y_p,q_p^2)$ 
obtained after integrating the $\nu_\mu$ and $\bar{\nu}_e$ momenta which is defined as
\begin{equation}
\label{Eq:diff-decay}
\overline{\Gamma}(x_p,y_p,q_p^2)=\frac{1}{\Gamma_\mu}\cdot\frac{d^3\Gamma}{dq_p^2
\,dx_p \,dy_p},
\end{equation}
where, $\Gamma_\mu$ is the total decay width of muon.
In terms of these new normalized variables, the phase space for this process is 
bounded by three surfaces: 
$q_p^2=0$, $x_p=1/2$ and $(q_p^4-q_p^2+x_p^2-y_p^2)=0$. It is easily seen 
from Eq.~\eqref{costheta}, the plane $x_p=1/2$ corresponds to $\theta=0^\circ$ 
and the curved surface $(q_p^4-q_p^2+x_p^2-y_p^2)=0$ signifies 
$\theta=180^\circ$. 
The physical region in $q_p^2$, $x_p$ and $y_p$ parameter space is given by,
\begin{gather}
q_p\sqrt{1-q_p^2}\leq x_p\leq\frac{1}{2},\nn\\
|y_p|\leq (\frac{1}{2}-q_p^2),\nn\\
(q_p^4-q_p^2+x_p^2-y_p^2) \geq 0,\label{phys-reg}\\
0\leq q_p^2\leq\frac{1}{2}.\nn
\end{gather}
Form Eq.~\eqref{param} and Eq.~\eqref{phys-reg}, it is clear that both $q_p^2$
and $x_p$  are positive valued functions whereas $y_p$ can have a positive value
or a negative value and the physical region allows $y_p$ to have a range
symmetric about $y_p=0$. So, if $(x_p,y_p,q_p^2)$ be a point inside physical
region, $(x_p,-y_p,q_p^2)$ will also lie inside the allowed region. This
motivates us to investigate the properties of odd and even part of
$\overline{\Gamma}(x_p,y_p,q_p^2)$ under the variable $y_p$. In the next 
section 
(Sec.~\ref{Sec:obs}) we construct such an observable  
as the ratio of odd part in $y_p$ divided by even part in $y_p$ of
$\overline{\Gamma}(x_p,y_p,q_p^2)$ and demonstrate its heightened sensitivity 
to 
$\eta_\gamma$.

\section{Observable and asymmetry}
\label{Sec:obs}

The `odd' and `even' part
$\overline{\Gamma}_o\,(x_p,y_p,q_p^2)$ and 
$\overline{\Gamma}_e\,(x_p,y_p,q_p^2)$, respectively, of the normalized
differential decay rate (Eq.~\eqref{Eq:diff-decay}) with respect to $y_p$ are 
defined as
\begin{align}
\label{eq:gammaO}
\overline{\Gamma}_o\,(x_p,y_p,q_p^2)&=\frac{1}{2}\Big[\overline{\Gamma}(x_p,y_p,q_p^2)-
\overline{\Gamma}(x_p,-y_p,q_p^2)\Big]\nn\\
&\approx F_o (x_p,y_p,q_p^2) + \eta_\gamma\,G_o 
(x_p,y_p,q_p^2),\\
\overline{\Gamma}_e\,(x_p,y_p,q_p^2)&=\frac{1}{2}\Big[\overline{\Gamma}(x_p,y_p,q_p^2)+
\overline{\Gamma}(x_p,-y_p,q_p^2)\Big]\nn\\
&\approx F_e (x_p,y_p,q_p^2) + \eta_\gamma\,G_e 
(x_p,y_p,q_p^2),
\end{align}
where the small $\eta_\gamma^2$ terms are ignored.


As we have obtained $\overline{\Gamma}(x_p,y_p,q_p^2)$ by integrating a 
positive
valued function $|\mathcal{M}|^2$, it is obvious that both
$\overline{\Gamma}(x_p,y_p,q_p^2)$ and $\overline{\Gamma}(x_p,-y_p,q_p^2)$ will 
be 
positive.
Hence, $\overline{\Gamma}_e\,(x_p,y_p,q_p^2)$, which is proportional to the sum 
of
$\overline{\Gamma}(x_p,y_p,q_p^2)$ and $\overline{\Gamma}(x_p,-y_p,q_p^2)$, as 
well as
${F}_e (x_p,y_p,q_p^2)$, which is $\eta_\gamma\to 0$ limit of
$\overline{\Gamma}_e\,(x_p,y_p,q_p^2)$, will always be greater than or equal to 
zero 
inside the physical region. On the other hand,
$\overline{\Gamma}_o\,(x_p,y_p,q_p^2)$, which is proportional to subtraction of 
two
positive quantities, as well as  ${F}_o (x_p,y_p,q_p^2)$, which is
$\eta_\gamma\to 0$ limit of $\overline{\Gamma}_o\,(x_p,y_p,q_p^2)$, could be 
positive, zero
or negative inside the allowed region.

We now define an observable, ${R}_\eta$, as 
\begin{equation}
{R}_\eta (x_p,y_p,q_p^2) 
=\frac{\overline{\Gamma}_o}{\overline{\Gamma}_e}\approx\frac{{F}_o}{{F}_e}
\bigg[1+\eta_\gamma\,
\Big(\frac{{G}_o}{{F}_o}-\frac{{G}_e}{{F}_e}\Big)\bigg]
\end{equation}
and  the asymmetry, $A_\eta (x_p,y_p,q_p^2)$, in ${R}_\eta$ as 
\begin{equation}
\label{asym}
A_\eta 
(x_p,y_p,q_p^2)=\Big(\frac{{R}_\eta}{{R}_{\rm SM}}-1\Big)
\approx\eta_\gamma\,
\Big(\frac{{G}_o}{{F}_o}-\frac{{G}_e}{{F}_e}\Big)
\end{equation}
where,
\begin{equation*}
{R}_{\rm 
SM}=\frac{\overline{\Gamma}_o}{\overline{\Gamma}_e}\bigg|_{\eta_\gamma=0}=\frac{{F}_o}{{F}_e}.
\end{equation*} 
Since, ${F}_o$ and ${G}_o$ are the zeroth order and first order terms
respectively in the expansion of the odd part of 
$\overline{\Gamma}(x_p,y_p,q_p^2)$ 
with
respect to $\eta_\gamma$ (see Eq.~\eqref{eq:gammaO}), both of them are expected 
to be proportional to odd
powers of $y_p$, rendering the ratio $({G}_o/{F}_o)$ to be finite at $y_p=0$.

We will now show that ${F}_o$ i.e. the odd part of SM,  has a zero for this 
mode for all $q_p^2$. For simplicity, to describe the situation mathematically, 
we consider
only the dominant contributions arising from the first and second Feynman 
diagrams in Fig.~\ref{feyn}. Retaining only relevant terms upto 
$\mathcal{O}(1/m_W^4)$,
we can write,
\begin{equation}
{F}_o\propto y_p\;h(x_p,y_p,q_p^2)\;f(x_p,y_p,q_p^2)
\end{equation}
where,
\begin{equation}
\label{eq:den}
h=\bigg[\frac{1+q_p^2}{(1-q_p^2)^5(1-2x_p)\{(1-q_p^2-x_p)^2-y_p^2\}^2}\bigg],
\end{equation}
\begin{equation}
\begin{split}
f=&\bigg[7\,q_p^8-4(4-x_p)\,q_p^6+(11-4x_p+6x_p^2-6y_p^2)\\
& q_p^4-2\,q_p^2\,(1-x_p+8x_p^2-6x_p^3-4y_p^2+2x_p y_p^2)\\
&+3x_p^4-12x_p^3+x_p^2(11-2y_p^2)-x_p(2-4y_p^2)\\
&-y_p^2(3+y_p^2)\bigg].
\end{split}
\end{equation}
As can be seen from the inequalities in Eq.~\eqref{phys-reg}, $h(x_p,y_p,q_p^2)$
is always positive inside the physical region. Hence, the deciding factor on 
the sign of $F_o$ is only $f(x_p,y_p,q_p^2)$. Now, on $x_p=1/2$ surface, we have
\begin{equation*}
f\Big(\frac{1}{2},y_p,q_p^2\Big)=\frac{7}{16}(1-2q_p^2)^4-
\frac{3}{2}(1-2q_p^2)^2\,y_p^2-y_p^4,
\end{equation*}
which after using the upper limit of $|y_p|$ from Eq.~\eqref{phys-reg}, implies 
that 
\begin{align}
&f\Big(\frac{1}{2},y_p,q_p^2\Big)\geq 0,\\
\implies~&{F}_o \Big(\frac{1}{2},|y_p|,q_p^2\Big)\geq 0,\\
&{F}_o \Big(\frac{1}{2},-|y_p|,q_p^2\Big)\leq 0.
\end{align}
Similarly for any point on the curved surface $(q_p^4-q_p^2+x_p^2-y_p^2)=0$ 
denoted as $C$, we have  $y_p^2=(q_p^4-q_p^2+x_p^2)$ and hence,
\begin{equation}
f(x_p,y_p,q_p^2)\Big|_C=(1-q_p^2)(1-2x_p)^2(q_p^2-2x_p).
\end{equation}
On using the limits of $x_p$ and $q_p^2$ from Eq.~\eqref{phys-reg}, it can 
easily be shown that
\begin{align}
&f(x_p,y_p,q_p^2)\Big|_C\leq 0,\\
\implies~&{F}_o (x_p,|y_p|,q_p^2)\Big|_C\leq 0,\\
&{F}_o (x_p,-|y_p|,q_p^2)\Big|_C\geq 0.
\end{align}
\begin{figure}[th!]
\includegraphics[scale=0.2]{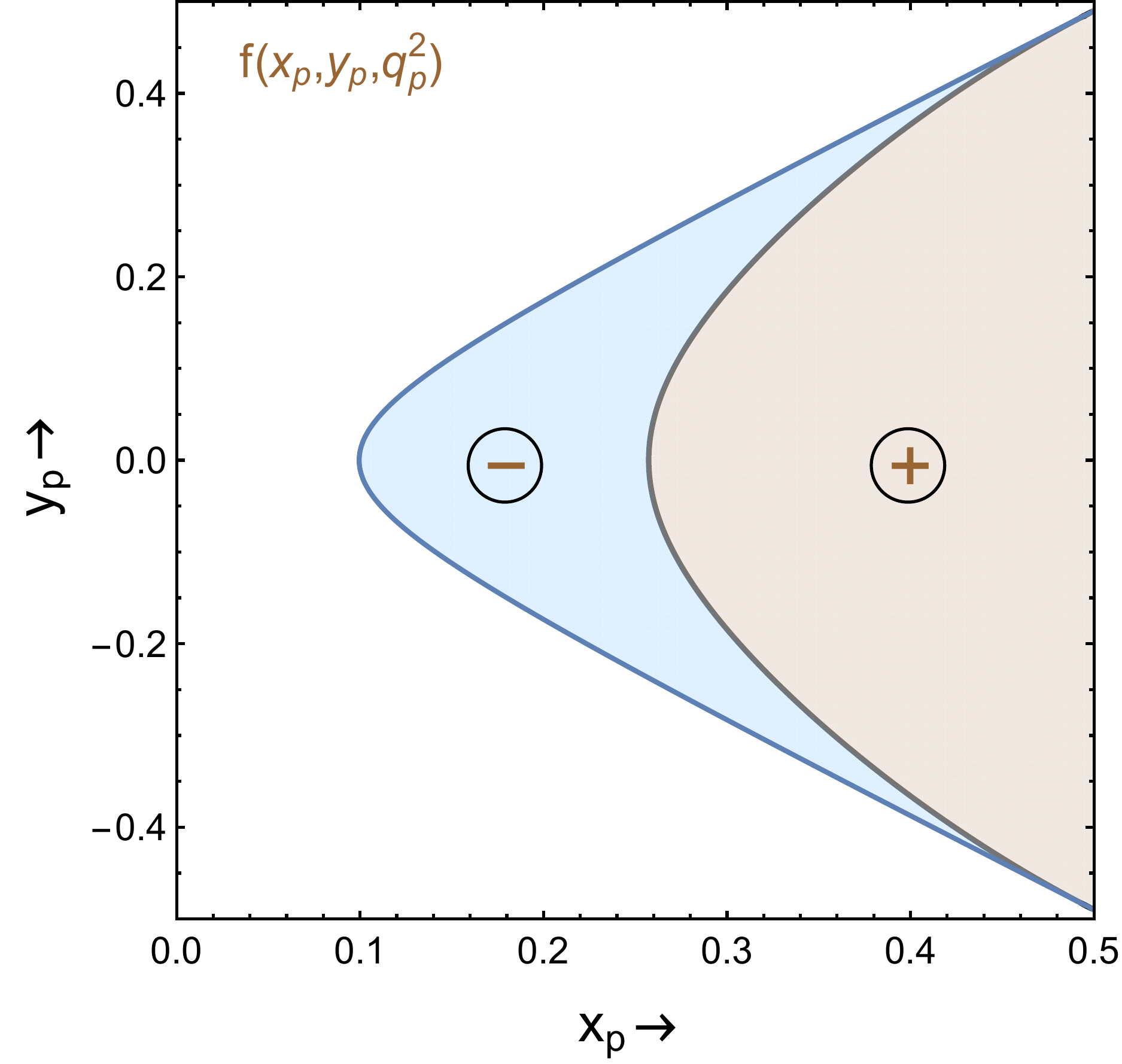}
\includegraphics[scale=0.2]{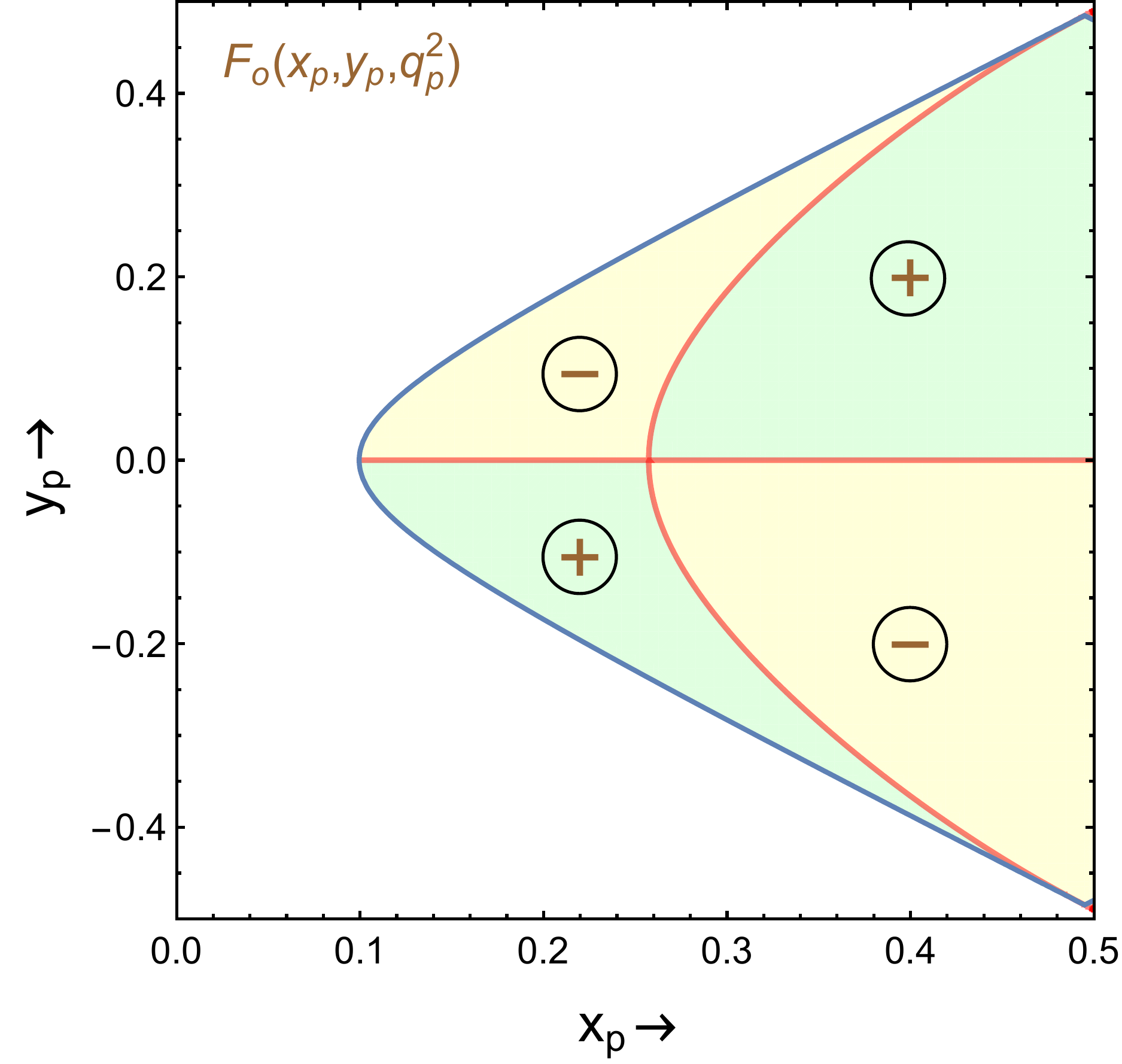}
\caption{The variations of functions $f(x_p,y_p,q_p^2)$ and ${F}_o(x_p,y_p,q_p^2)$ are shown in $x_p-y_p$ plane in left and right panel, respectively, where $q_p^2=0.01$.  The blue 
line in both the panels indicates one boundary of phase space with 
$\cos\theta=-1$ or $(q_p^4-q_p^2+x_p^2-y_p^2)=0$. In the left panel, the blue region signifies negative valued $f(x_p,y_p,q_p^2)$, 
the brown region symbolizes positive valued $f(x_p,y_p,q_p^2)$ and 
the black curve indicates $f(x_p,y_p,q_p^2)=0$. In the right panel, 
the yellow region signifies negative valued 
${F}_o(x_p,y_p,q_p^2)$, the green region symbolizes 
positive valued ${F}_o(x_p,y_p,q_p^2)$ and the red curve 
indicates ${F}_o(x_p,y_p,q_p^2)=0$. }
\label{f-F_o}
\end{figure}
We have concluded that $f(x_p,y_p,q_p^2)<0$ along the curve $C$ and
$f(x_p,y_p,q_p^2)>0$ at the other boundary surface $x_p=1/2$. It is obvious
therefore that there must be at least one surface within the allowed phase space
region where $f(x_p,y_p,q_p^2)=0$. In the first plot of Fig.~\ref{f-F_o}, the
blue region signifies $f(x_p,y_p,q_p^2)<0$ and the brown region
symbolizes $f(x_p,y_p,q_p^2)>0$ whereas the black curve indicates
$f(x_p,y_p,q_p^2)=0$. In the second plot  of Fig.~\ref{f-F_o}, the yellow region
signifies ${F}_o(x_p,y_p,q_p^2)<0$ and the green region symbolizes
 ${F}_o(x_p,y_p,q_p^2)>0$ while the red curve indicates
${F}_o(x_p,y_p,q_p^2)=0$. 

\begin{figure*}[t!]
\includegraphics[scale=0.17]{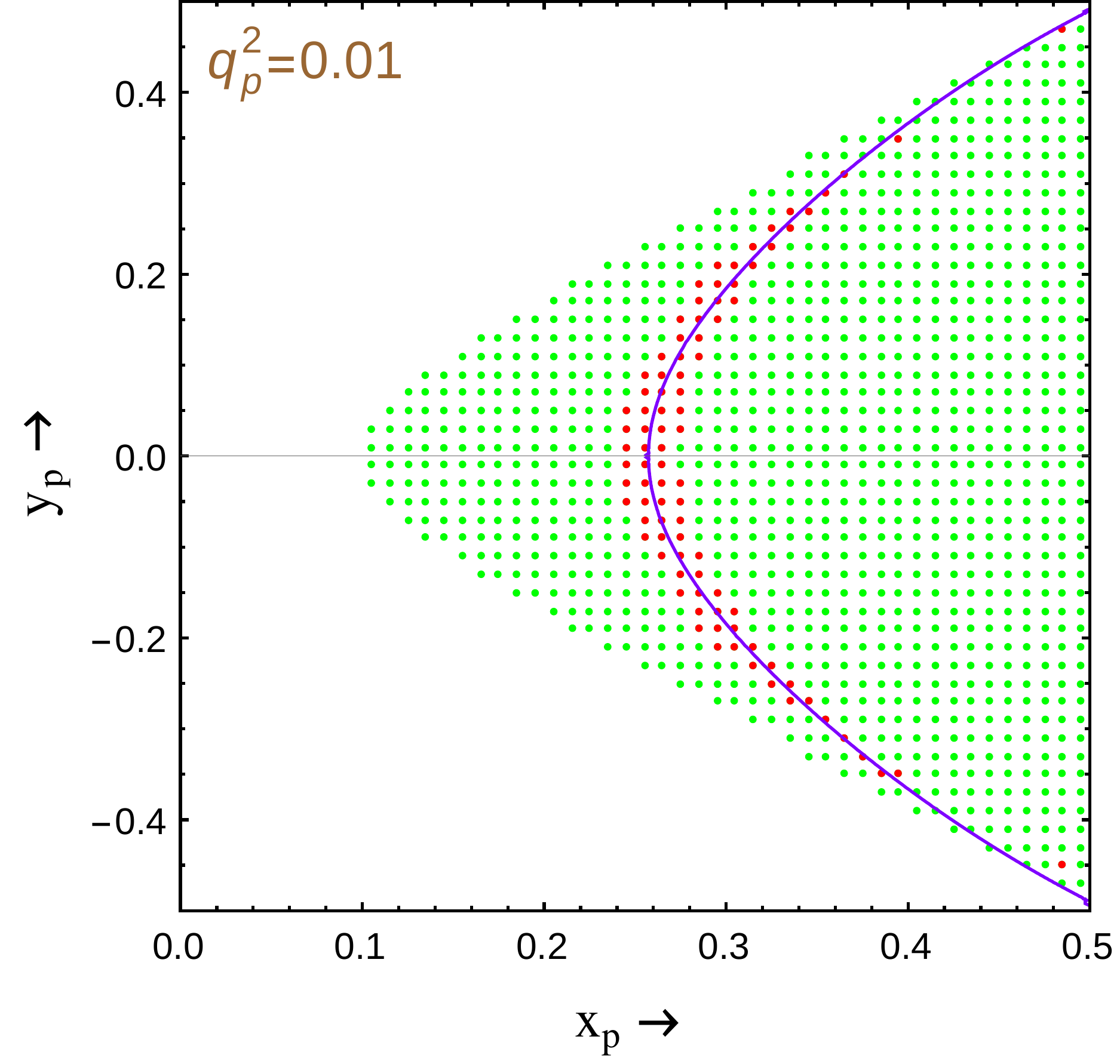}
\includegraphics[scale=0.17]{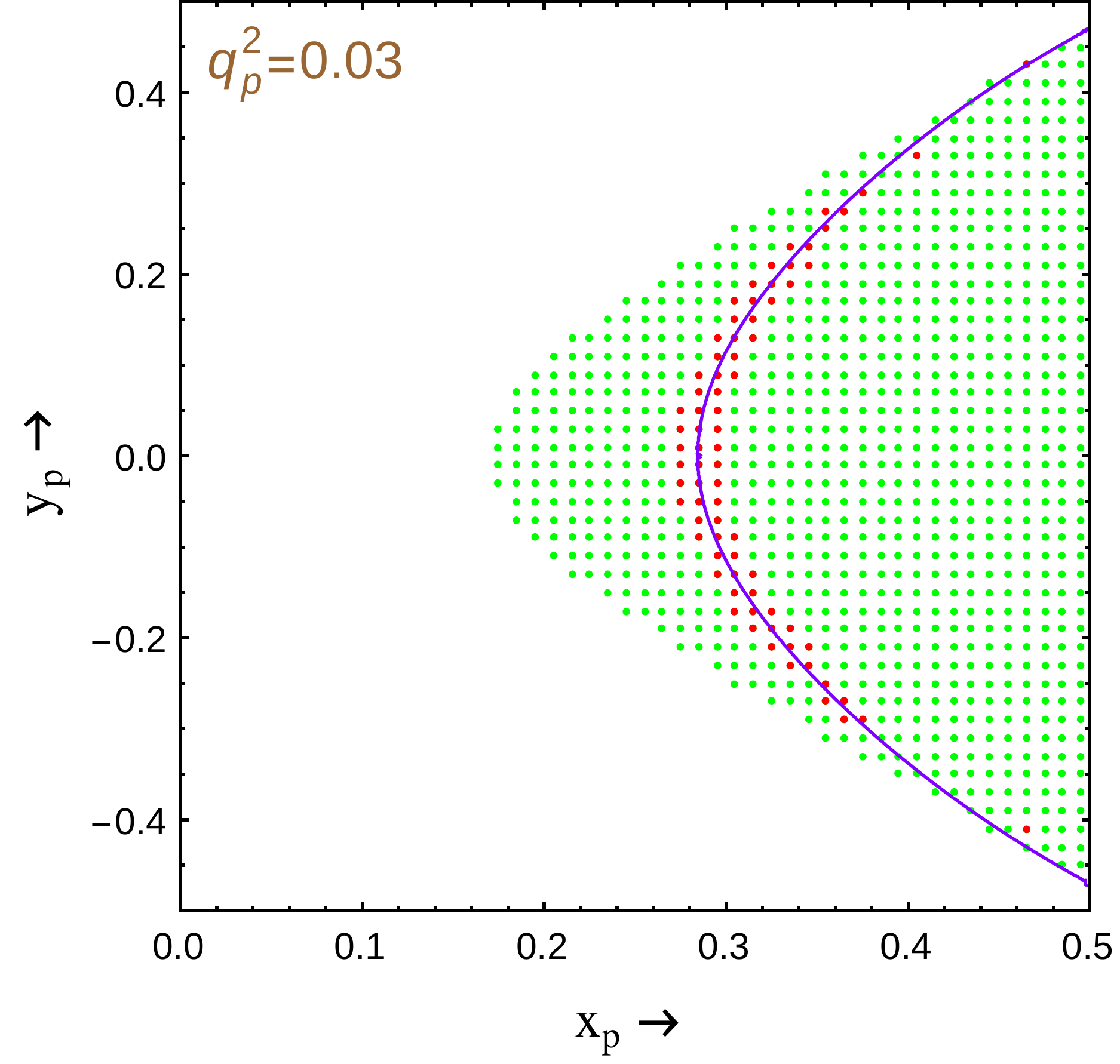}
\includegraphics[scale=0.17]{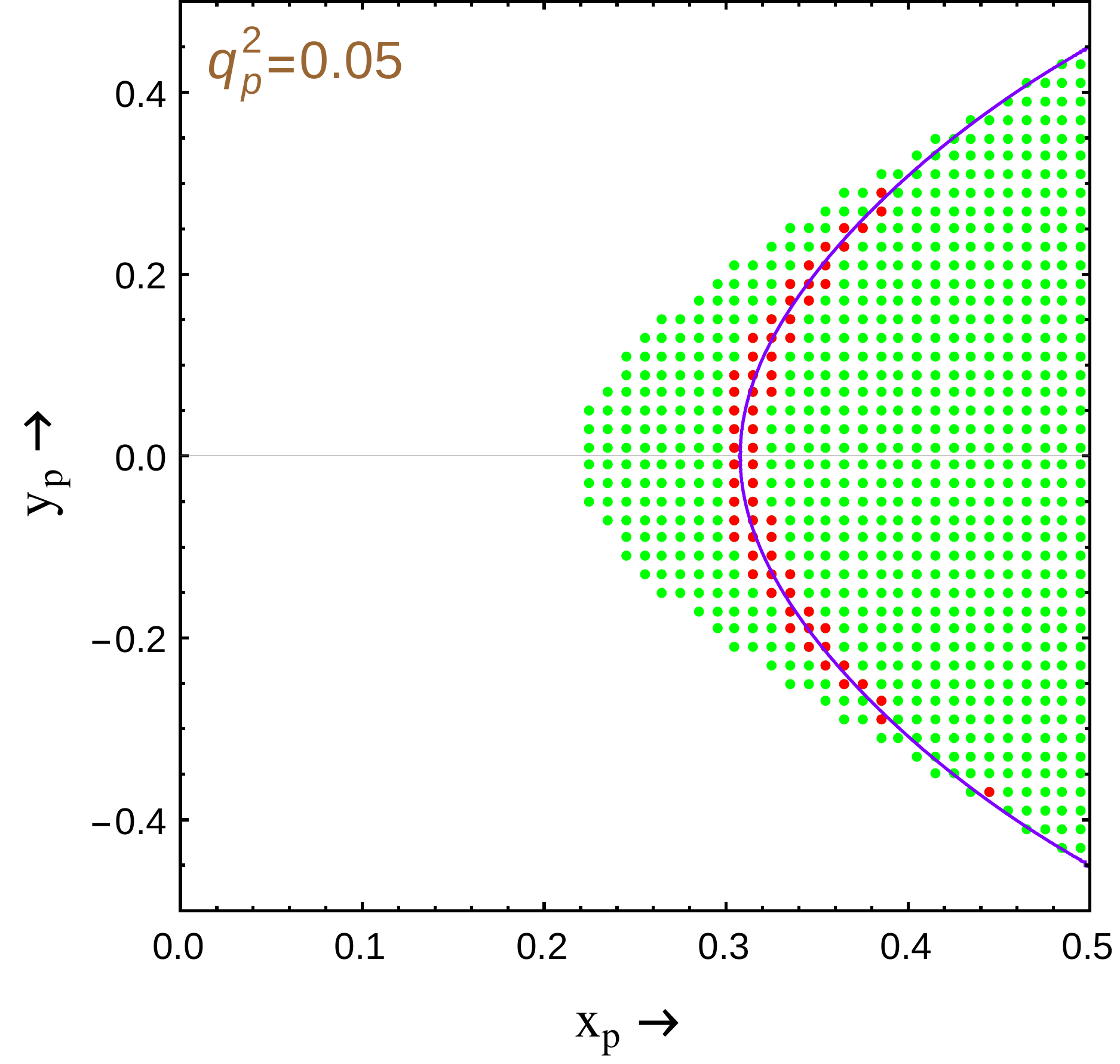}
\includegraphics[scale=0.17]{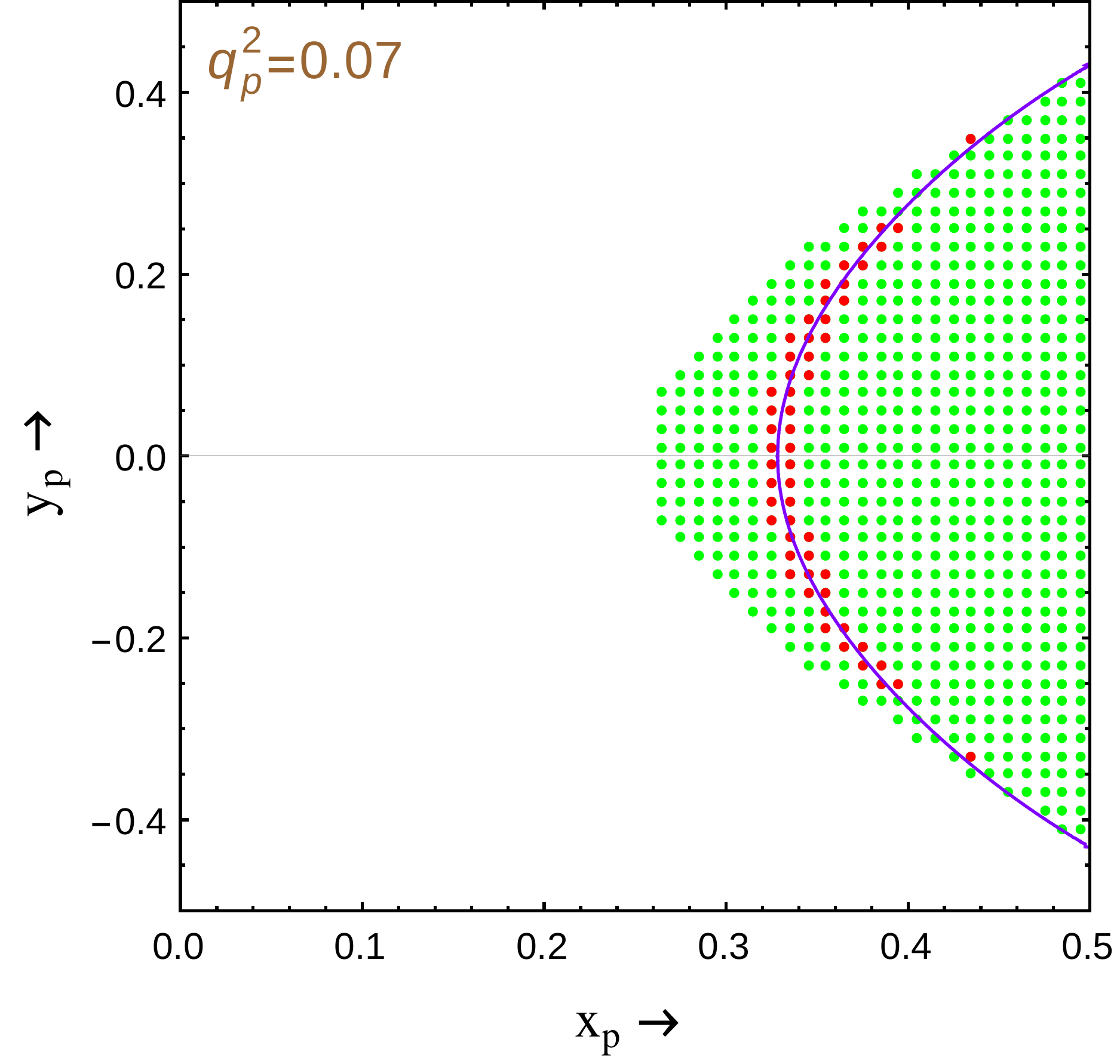}
\includegraphics[scale=0.17]{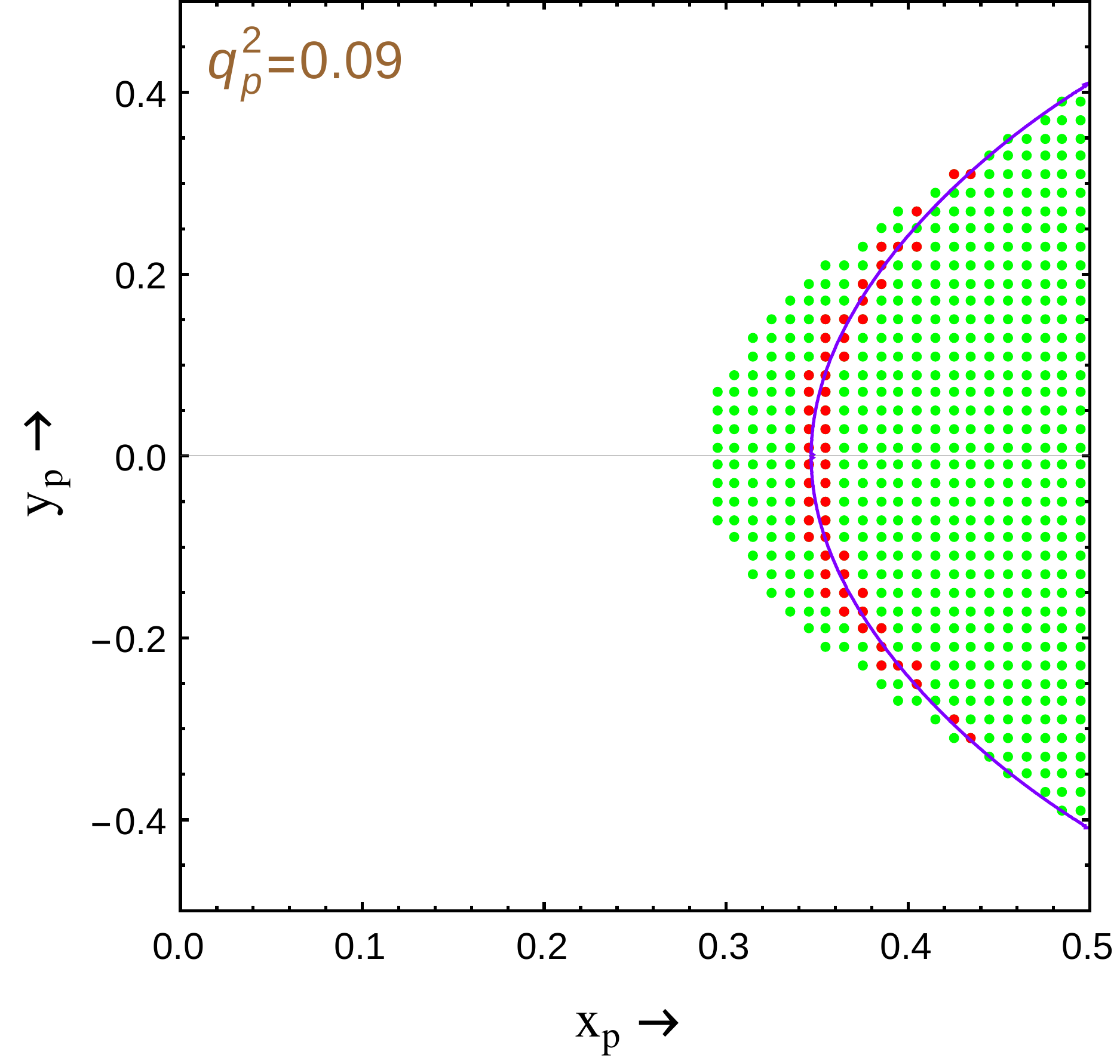}
\includegraphics[scale=0.17]{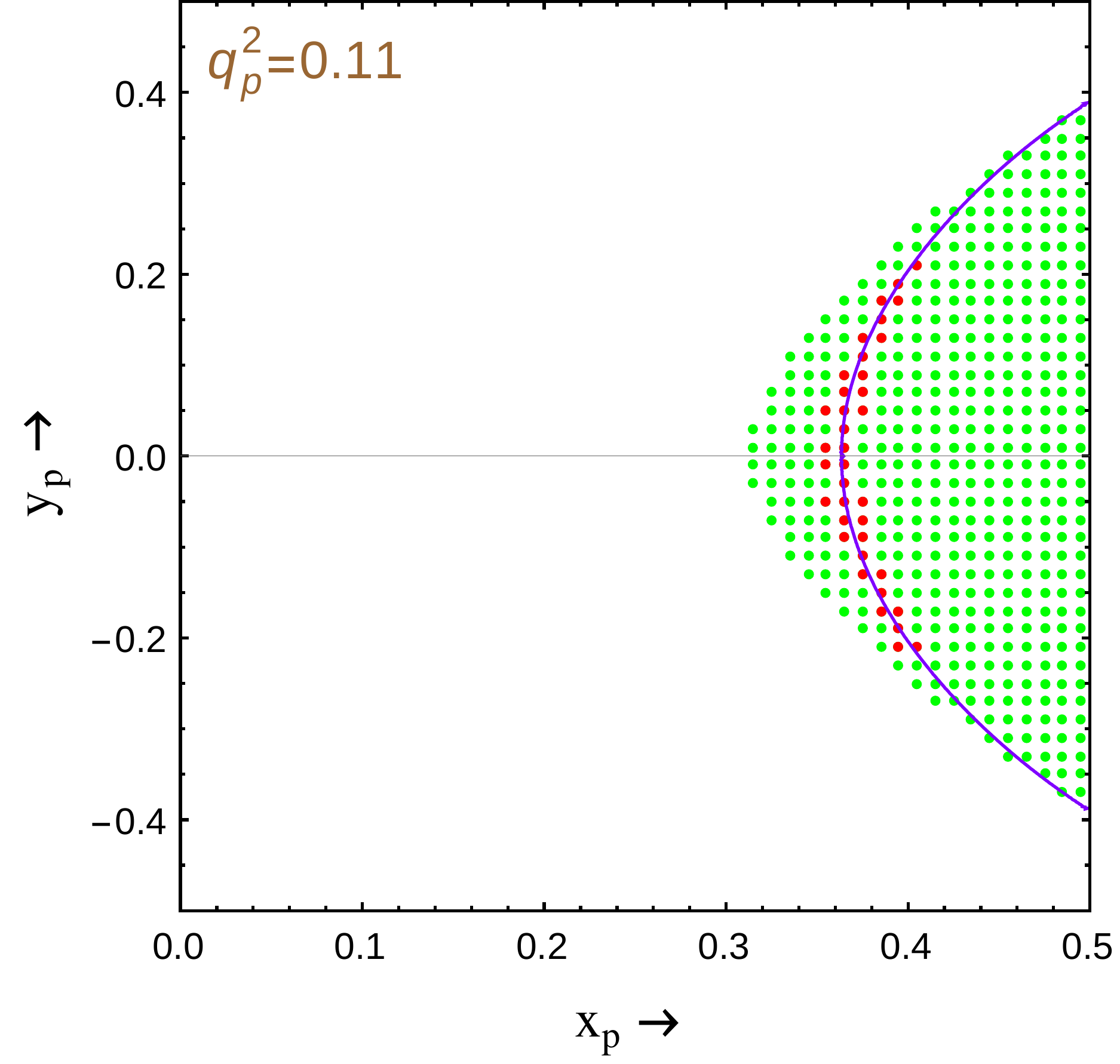}
\includegraphics[scale=0.17]{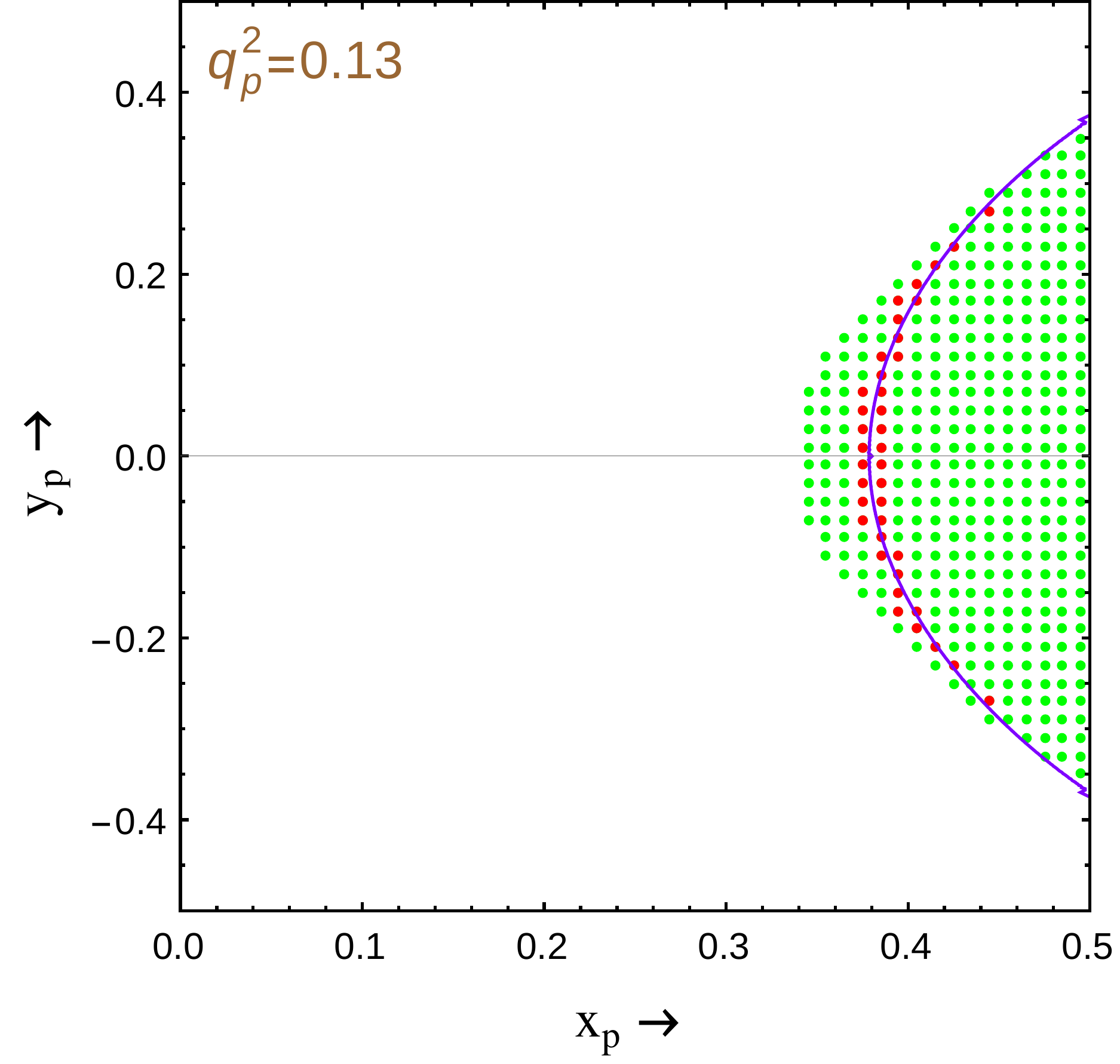}
\includegraphics[scale=0.17]{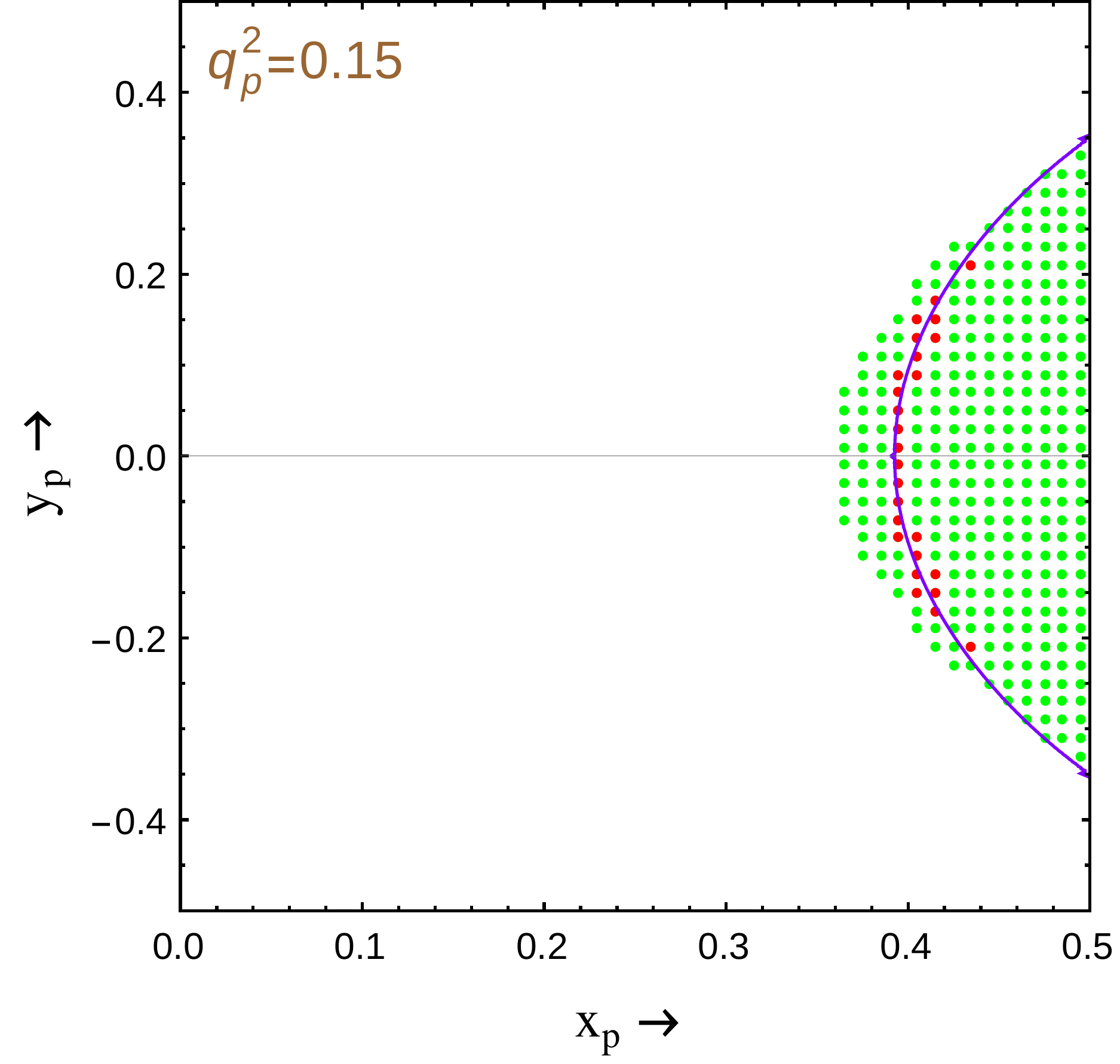}
\includegraphics[scale=0.17]{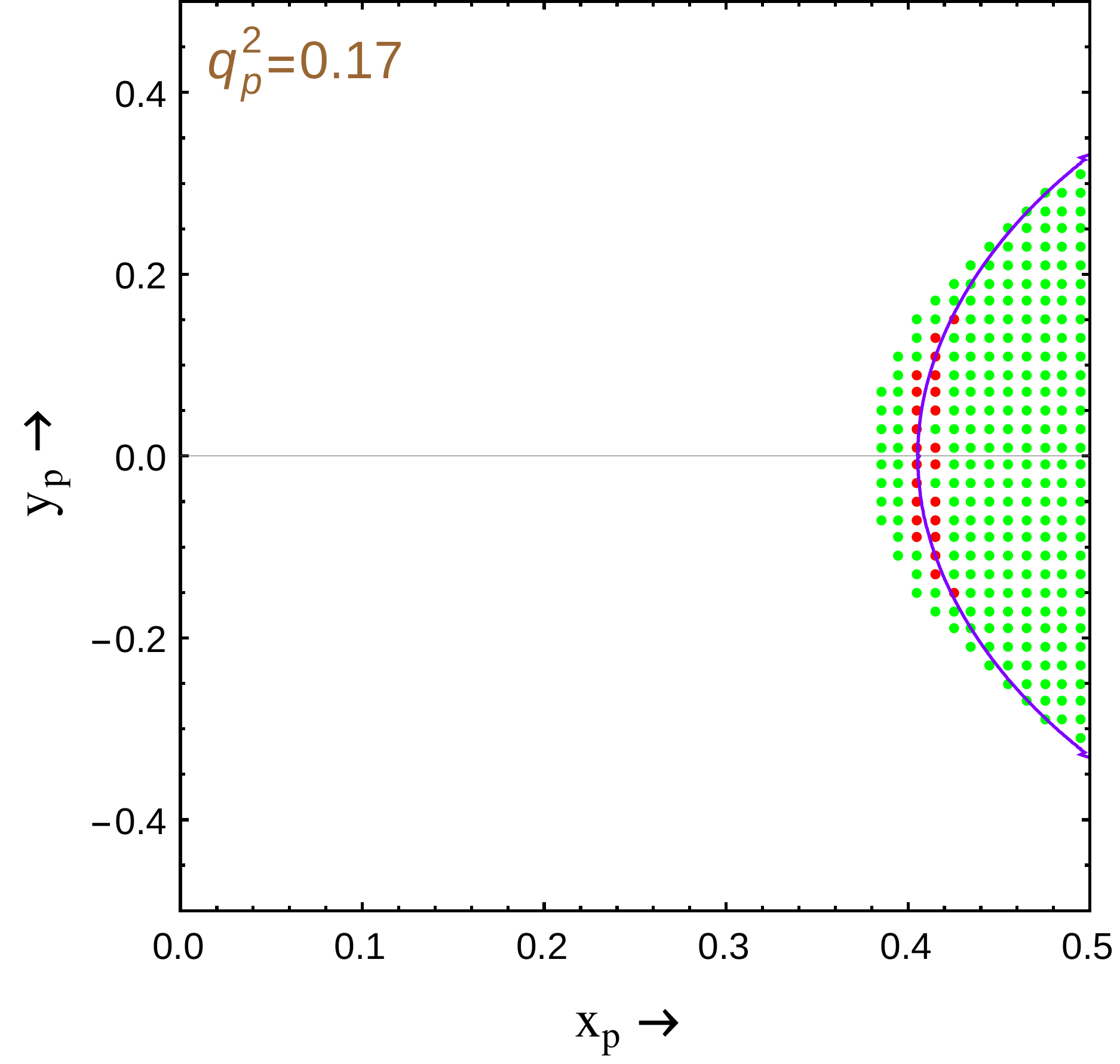}
\includegraphics[scale=0.17]{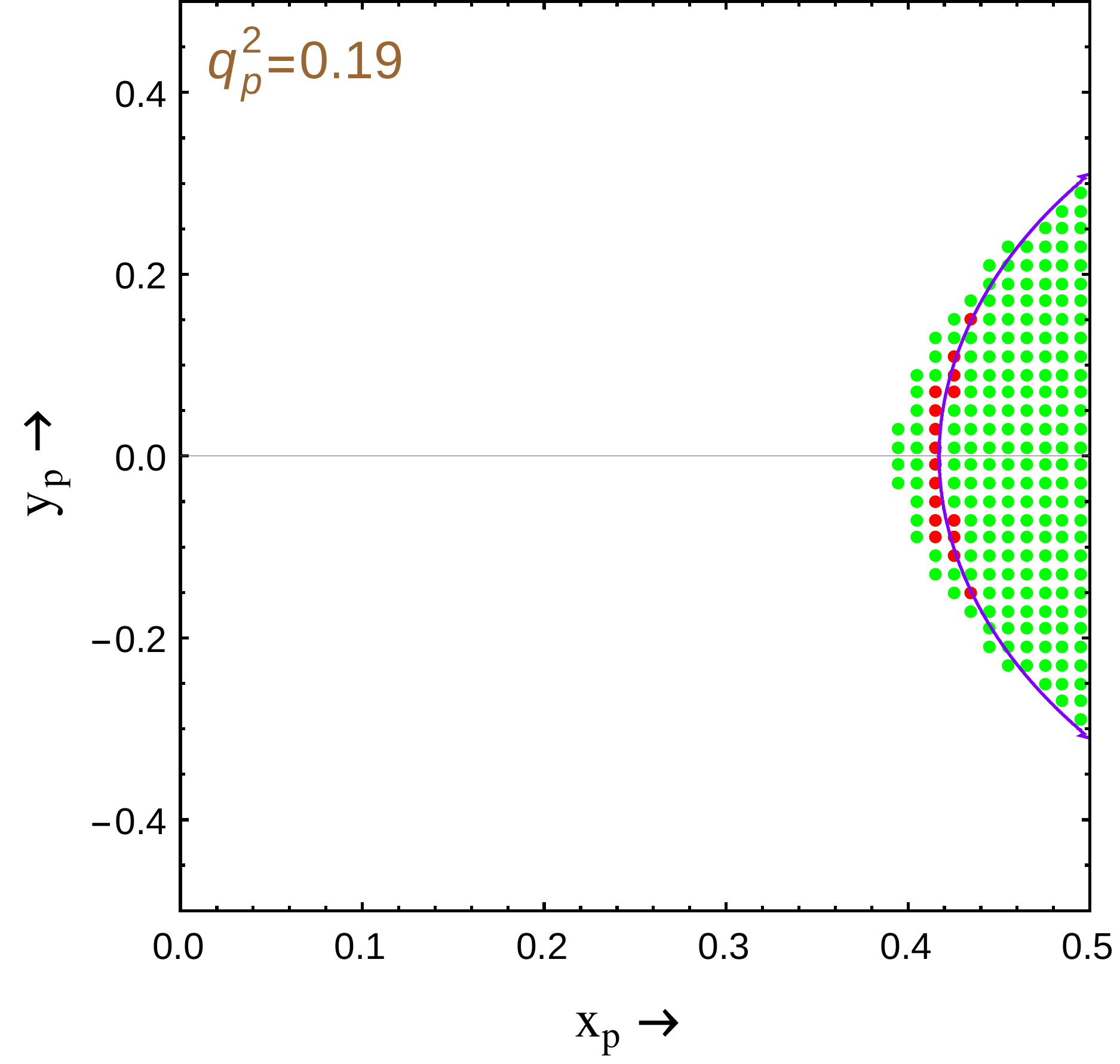}
\caption{The variation of $F_o(x_p,y_p,q_p^2)$ for different $q_p^2$ in the $x_p$-$y_p$ 
plane.   Each green dot represents a bin 
according to experimental 
resolution of photon energy, electron energy and angle between 
them. The red dots stand for the bins having  
$\delta|A_\eta|/|A_\eta|\leq 10$ in that bin. 
The purple curve signifies ${F}_o=0$ in different $q_p^2$ 
plane. Our numerical analysis includes the bins corresponding to the red dots only. This 
results in an optimal sensitivity to $\eta_\gamma$.}
\label{fig:eta-reg}
\end{figure*}

The odd ($\overline{\Gamma}_o$) and even ($\overline{\Gamma}_e$) parts of
differential rate  as well as the four functions $F_o$, $F_e$, $G_o$, $G_e$
contain soft collinear divergences arising due to $E_\gamma=0$ or $\cos\theta=1$
and divergence due to vanishing $E_e$ if $m_e$ is ignored. It is obvious  form
Eq.~\eqref{Egamma} that soft photon dominate in the region corresponding to
$(x_p+y_p)\approx(1-q_p^2)$ , which implies $(x_p+y_p)$ is close to its maximum
value. Hence, events with small photon energy lie at the top corner in
Fig.~\ref{f-F_o} where the blue  curve  meets $x_p=\nicefrac{1}{2}$ line.
Similarly, one can see from Eq.~\eqref{Ee} that small electron energy implies
$(x_p-y_p)\approx(1-q_p^2)$ and these events lie at the bottom corner in
Fig.~\ref{f-F_o} where the blue curve  meets $x_p=\nicefrac{1}{2}$ line. For any
value of $q_p^2$, the collinear divergence occurs along $x_p=\nicefrac{1}{2}$
line as can easily be seen from Eq.~\eqref{costheta}. These singularities are
evident from Eq.~\eqref{eq:den} and occur in each of $\overline{\Gamma}$,
$\overline{\Gamma}_o$, $\overline{\Gamma}_e$ as well as the four functions
$F_o$, $F_e$, $G_o$, $G_e$. It is  only in these regions that an expansion in
powers of $m_e/m_\mu$ is not valid; the electron mass needs to be retained and
ignoring it alters the differential decay rates. To deal with the
$x_p=\nicefrac{1}{2}$ collinear singularity we choose an appropriate cut on
$x_p$ which is also necessitated by experimental resolution. It can be seen from
Eq.~\eqref{asym}, however, that with in SM, $A_\eta$  is finite and zero, even
in the regions plagued by collinear soft photon singularities and the ones that
arise due to neglect of $m_e$. Note, that in $A_\eta$ the $h$-function~in
Eq.~\eqref{eq:den} carrying the singular denominator cancels. The zero observed
in ${F}_o$ and the consequent singularity in the asymmetry $A_\eta$ has nothing
to do with the well know collinear soft photon and $m_e\to 0$ singularities. The
zero observed in ${F}_o$ is genuine and looks like an apparent exchange symmetry
between $E_e$ and $E_\gamma$ only for the appropriately chosen parameters, $x_p$
and $y_p$ (or $x_n$ and $y_n$ defined in Eq.~\eqref{eq:newxy}) with $m_e$
retained.

We have explicitly demonstrated that there exists a surface (besides $y_p=0$
plane) where ${F}_o(x_p,y_p,q_p^2)=0$; we refer to this surface corresponding to
the \emph{`new type of zero'} as ``null-surface''. This means that at each point
on this surface the differential decay rate $\overline{\Gamma}(x_p,y_p,q_p^2)$
remains unaltered if we interchange the energies of photon and electron. Hence,
$A_\eta (x_p,y_p,q_p^2)$ diverges on null-surface for any non-zero value of
$\eta_\gamma$ and becomes zero everywhere in the phase space for $\eta_\gamma$
being zero. The null-surface divides the phase space into two regions, one where
$A_\eta$ is positive and the other where $A_\eta$ is negative. For
$\eta_\gamma>0$,  $A_\eta<0$ for  $x_p$ values smaller than the values indicated
by the null-surface, whereas, $A_\eta>0$ for  $x_p$ values larger than the
values indicated by the null-surface. However, if $\eta_\gamma<0$, an opposite
behaviour in the signs of $A_\eta$ is indicated. This feature can be used to
determine the sign of $\eta_\gamma$. To measure the value of $\eta_\gamma$
experimentally, one must average $A_\eta$ over specified regions of phase space
where it could be positive or negative. Such averages are necessitated by the
experimental resolutions for $q_p^2$, $x_p$ and $y_p$ and  will in general
reduce the asymmetry. Hence, it is convenient to use $|A_\eta|$ as the
asymmetry.

 In the next section (Sec.~\ref{Sec:simulation}) we probe the
feasibility to measure $\eta_\gamma$ using the asymmetry obtained in this
section.

\section{Simulation and analysis}
\label{Sec:simulation}

In order to study the sensitivity of muon radiative decay mode we need to
include the resolutions for energy of photon, energy of electron and the angle
between them. We take them to be $2\%$, $0.5\%$ and $10$ Milli-radian,
respectively~\cite{Kuno}. As can be seen from Eq.~\eqref{Ee}-\eqref{costheta},
the resolutions for $x_p$, $y_p$ and $q_p^2$ will also vary at different point
in phase space due to the functional form of these parameters. We begin by
evaluating the resolutions for $x_p$, $y_p$ and $q_p^2$ for the entire allowed
phase space. We find that the resolutions for $x_p$ $y_p$ and $q_p^2$ are always
less than  $0.01$, $0.02$ and $0.02$ respectively. For simplicity, in our
simulation, we take the worst possible scenario and assume constant resolutions
for each of $x_p$, $y_p$ and $q_p^2$, corresponding to their largest value of
$0.01$, $0.02$ and $0.02$ respectively throughout the entire allowed phase
space, which allows us to choose equal size bins. Hence, the phase space region
$0\leq q_p^2\leq\nicefrac{1}{2}$, $0\leq x_p\leq\nicefrac{1}{2}$,
$\nicefrac{-1}{2}\leq y_p\leq\nicefrac{1}{2}$ is divided into $25$ bins in 
$q_p^2$ and $50$ bins in both $x_p$ and $y_p$ -- all equal in size. Among these 
bins, only 6378 number of bins
lie inside the physical phase space region. We next estimate the systematic and 
statistical error for $|A_\eta|$ in each of these bins, assuming
$\eta_\gamma=0.01$.

To find the systematic error in $|A_\eta|$ for a particular $i$-th bin,  we
evaluate it at $62,500$  equally spaced points in that bin to estimate
$|A_\eta|_i^j$ where $j$ is the index of a point inside the $i$-th bin. However,
for the bins near to the boundary of phase space, all of these points will not
be inside the physical region and hence, we denote the number of physical points
inside $i$-th bin as $n_i^{}$. We now, calculate the average of $|A_\eta|_i^j$
inside a bin, i.e.
$$\braket{|A_\eta|_i}=\frac{1}{n_i^{}}\sum_j |A_\eta|_i^j,$$ and take this as 
the asymmetry of that bin. Then we take the systematic error as the average
deviation of $|A_\eta|_i^j$, i.e. $$\sigma_i^{\text{sys}}=\frac{1}{n_i^{}}\sum_j
\big|\braket{|A_\eta|_i}-|A_\eta|_i^j\big|.$$  
Ideally the errors can and should have been calculated using a standard
Monte-Carlo technique with more number of sample points. The approach followed
in this paper is to express the integral as a Riemann sum only  for simplicity.


The statistical error for $|A_\eta|$ in each bin is also estimated by averaging
it at the same $62,500$ equally spaced points. Note that, while $A_\eta$ is
divergent on the null-surface the average value of $|A_\eta|$ for the $i$-th
bin, i.e. $\braket{|A_\eta|_i}$, estimated from Monte Carlo studies is never
larger than $10^{-6}$ for any bin. Hence,
$$\sigma_i^{\text{sta}}=\sqrt{\frac{1-\braket{|A_\eta|_i}^2}{N_i}}\approx 
\frac{1}{\sqrt{(N_{SM})_i}},$$ where $i$ is the 
index of the bins and $N_i$ represents the number of events inside $i$-th bin
which is almost the same as $(N_{SM})_i$ the number of SM events for the $i$-th
bin. We have also assumed that both $A_\eta$ and the effects of $\eta_\gamma$ on
$N_i$  are small and can be ignored. If this were not the case $N_i$ would
itself be sensitive to $\eta_\gamma$, contrary to our simulation results.
Hence, we simply take the statistical error for all practical purposes to be
that in the case of SM events.  The number of events in each bin is
calculated by taking total number of muons to be $10^{19}$. To avoid the 
singularities in the number of SM events for the bins near 
$x_p=\nicefrac{1}{2}$ plane, we ignore the bins with $0.49\le x_p\le 0.5$.

The total error in $|A_\eta|$ for any particular bin is then given by
$\delta|A_\eta|_i=\sqrt{(\sigma_i^{\text{sta}})^2+(\sigma_i^{\text{sys}})^2}$.
This error in $|A_\eta|$ will affect the measurement of $\eta_\gamma$. Using
Eq.~\eqref{asym}, we observe that the error in the measurement of $\eta_\gamma$ 
in
each bin as
\begin{equation}
\label{eq:delta-eta}
\Big|\frac{\delta\eta_\gamma}{\eta_\gamma}
\Big|_i=\frac{\delta|A_\eta|_i}{|A_\eta|_i}
\end{equation}
where, $|A_\eta|_i\equiv\braket{|A_\eta|_i}$ and we take the theoretical
function $\Big({G}_o/{{F}_o}-{G}_e/{{F}_e}\Big)$ to be free from experimental
uncertainties. It is obvious from Eq.~\eqref{eq:delta-eta}, that the highest
sensitivity is achieved in bins close to the null-surface where $|A_\eta|_i$ is
the largest. Hence, we consider only the region along the null-surface by
applying a cut ${\delta|A_\eta|_i}/{|A_\eta|_i}\leq 10$ to determine 
$\eta_\gamma$. 

In Fig. \ref{fig:eta-reg}, we depict the bins, which satisfy the above cut, 
with red
dots for different $q_p^2$ values, whereas, the green dots signify all the other
bins inside the physical region; the purple curve indicates the null-surface
where ${F}_o=0$ for the corresponding $q_p^2$ value. Including only the bins,
which satisfy the above cut,  for a simulated value of $\eta_\gamma=0.01$ (at 
one loop in SM, $|\eta_\gamma|\lesssim 0.015$), we estimate  an error of
$\delta\eta_\gamma=2.6\times10^{-3}$, implying a $3.9\sigma$ significance for
the measurement. %
A total of $10^{19}$ muons  are aimed for in the long term future. The
next-round of experiments are aiming at $10^{18}$ muons /year. This reduces the
sensitivity from $3.9\sigma$ to $1.4\sigma$. To appreciate the advantage of
radiative muon decays in measuring $WW\gamma$ vertex one needs to note that the
current global average of $\kappa_\gamma$ differs from unity only by
$0.4\sigma$. We note that the significance of the measured value of
$\eta_\gamma$ may in principle be improved by optimizing the chosen cut and
binning procedure. However, we refrain from such intricacies as our approach is
merely to present a proof of principle.

We have shown that the sensitivity to $\eta_\gamma$ arises due to the
vanishing of the odd differential decay rate in the standard model denoted by
$F_o$. The observed singularity in $A_\eta$  is unrelated to soft
photon and collinear singularities or the singularity arising due to neglect of
$m_e$ in calculations. The most sensitive region  to measure $\eta_\gamma$ is
where $A_\eta$ is large and obviously lies along the zero of $F_o$ as indicated
by Eq.~\eqref{asym}. The region around $F_o=0$ for which 
${\delta|A_\eta|_i}/{|A_\eta|_i}\leq 10$, is where a legitimate expansion
in powers of $m_e/m_\mu$ can be carried out and is distinct from the singular
regions in the differential decay rates where such an expansion cannot be done.
However, in order to verify the accuracy of sensitivity achievable in
$\eta_\gamma$ measurement the calculations have been redone by numerically
retaining $m_e$. We find that for the bins represented by red dots in 
Fig.~\ref{fig:eta-reg} the maximum correction in $\eta_\gamma$ is ${\cal 
O}(10^{-4})$, which is an order of magnitude smaller than the error in it,
$\delta\eta_\gamma=2.6\times10^{-3}$.

Finally, we discuss possible sources of inaccuracies in our
estimation of uncertainty. Higher order electroweak corrections to the process 
considered will
modify the decay rate and alter $F_o$. While higher order electroweak 
corrections have not been included in our analysis they have been worked out in 
detail \cite{rad. cor.}.  However, this is unlikely to affect our
analysis technique as we have selected bins to be included in estimating
$\eta_\gamma$ purely based on the criterion ${\delta|A_\eta|_i}/{|A_\eta|_i}\leq
10$ and not on the location and validity of the null-surface. A possible source
of uncertainty that we have ignored in our analysis is the assumption that the muon
decays at rest or with known four-momenta. While facilities that produce large
numbers of muons are designed to bring the muon to rest, a  fraction of them may
decay with a finite but unknown 4-momenta, rendering the exact measurement of
$q_p^2$ inaccurate. This effect can in-principle be considered by including
additional systematic errors  in $q_p^2$.

\section{Conclusion}
\label{Sec:conclusion}

In order to probe lepton flavor violating process $\mu\to e\gamma$, facilities
that produce large numbers of muons are being designed. We show that radiative
muon decay $\mu\to e \gamma\nu_\mu\bar{\nu}_e$ is a promising mode to probe loop
level corrections in the SM to the $C$ and $P$ conserving dimension four $WW\gamma$
vertex with good accuracy. The process has two missing neutrinos in the final
state and on integrating their momenta the partial differential decay rate
removes the well known radiation-amplitude-zero. We show, however, that  the
normalized differential decay rate, odd under the exchange of photon and
electron energies, does have a zero in the case of standard model (SM). This 
\emph{new
type of zero} had hitherto not been studied in literature. A suitably
constructed asymmetry using this fact enables a sensitive probe for  the
$WW\gamma$ vertex beyond the SM. The large number of muons produced keeps the
statistical error in control for a tiny part of the physical phase space,
enabling us to measure $\eta_\gamma=0.01$ with  $3.9\sigma$ significance. %
\begin{acknowledgments}
 We thank Yoshitaka Kuno, Marcin Chrz\k{a}szcz, Thomas G. Rizzo and Jernej F.
 Kamenik  for valuable suggestions and discussions. The work of RM is supported 
 in part by Grants No. FPA2014-53631-C2-1-P, FPA2017-84445-P and SEV-2014-0398 
 (AEI/ERDF, EU) and by PROMETEO/2017/053. 
 \end{acknowledgments}

\appendix
\section{Expressions with electron mass retained}
\label{Sec:Appendix}

	In presence of electron mass $m_e$, we have $s+t+u=q^2+m_\mu^2+m_e^2$ where 
	the Mandelstam variables are defined as: $(p_e+p_\gamma)^2=s$,  
	$(p_e+q)^2=t$ and
	$(p_\gamma+q)^2=u$. 
	The physical region is determined by the following inequalities 
	\cite{byckling}:
	\begin{eqnarray}
	& m_e^2\leq s \leq (m_\mu-\sqrt{q^2})^2,&\\
	& q^2\leq u\leq (m_\mu-m_e)^2,&\\
	&(m_e+\sqrt{q^2})^2\leq t\leq m_\mu^2,&\\
	& G[s,u,m_\mu^2,0,m_e^2,q^2]\leq 0.	\end{eqnarray}
	where 
	\begin{equation}
	G[x,y,z,u,v,w]=-\frac{1}{2}\begin{vmatrix}
	0 & 1 & 1& 1& 1\\
	1 & 0 & v & x& z\\
	1& v& 0& u& y\\
	1& x& u& 0& w\\
	1& z& y& w& 0
	\end{vmatrix}
	\end{equation}
	
	We define variables $x_n,y_n$ and $q_n^2$, which reduce to $x_p,y_p$ and 
	$q_p^2$ at $m_e\rightarrow 0$ limit,
in the following way:
\begin{equation}
\label{eq:newxy}
\begin{split}
x_n=\frac{t+u}{2(q^2+m_\mu^2+m_e^2)},\\
y_n=\frac{t-u+m_e^2}{2(q^2+m_\mu^2+m_e^2)},\\
q_n^2=\frac{q^2}{(q^2+m_\mu^2+m_e^2)},
\end{split}
\end{equation}
The energy of electron and photon are obtained from the above definitions as:
\small
\begin{eqnarray}
&\hspace{-7mm} E_e=\dsp\frac{(2m_\mu^2+m_e^2)(1-q_n^2-x_n+y_n)-m_e^2(x_n-y_n)}{4 
m_\mu(1-q_n^2)},\\
&\hspace{-7mm} E_\gamma=\dsp\frac{(2m_\mu^2+m_e^2)(1-q_n^2-x_n-y_n)-m_e^2(x_n+y_n)}{4 
m_\mu(1-q_n^2)}.
\end{eqnarray}
\normalsize
Under the replacement $y_n\rightarrow-y_n$ electron and photon energies get
exchanged and one separate the odd and even parts differetial decay rate as follows: 
\begin{equation}
\begin{aligned}
\hspace{-2mm}\overline{\Gamma}_o\,(x_n,y_n,q_n^2)&=\frac{1}{2}
\Big[\overline{\Gamma}(x_n,y_n,q_n^2)-
\overline{\Gamma}(x_n,-y_n,q_n^2)\Big]\\
\hspace{-2mm}\overline{\Gamma}_e\,(x_n,y_n,q_n^2)&=\frac{1}{2}
\Big[\overline{\Gamma}(x_n,y_n,q_n^2)+
\overline{\Gamma}(x_n,-y_n,q_n^2)\Big]
\end{aligned}
\end{equation}
The $h$-function in Eq.~\eqref{eq:den} containing singular denominator, now, becomes
\begin{equation}
h\propto\frac{1}{E_e^2\,E_\gamma^2\,(m_\mu^2 (1-2 x_n)+m_e^2 (q_n^2-2 x_n))}.
\end{equation}

The region around $F_o=0$ which are denoted by red dots in Fig.~\ref{fig:eta-reg}, a legitimate expansion 
in powers of $(m_e/m_\mu)$ for the expressions of $\overline\Gamma_o$ and $\overline\Gamma_e$ can be carried out in the following way:
\small
\begin{align}
\hspace{-3mm}\overline{\Gamma}_o
&\approx (F_o +(m_e/m_\mu)^2\, \delta F_o) + \eta_\gamma(\,G_o +(m_e/m_\mu)^2\, \delta G_o)\\
\hspace{-2mm}\overline{\Gamma}_e
&\approx (F_e +(m_e/m_\mu)^2\, \delta F_e) + \eta_\gamma(\,G_e +(m_e/m_\mu)^2\, \delta G_e)
\end{align}
\normalsize
where the small $\eta_\gamma^2$ terms are ignored. Here, $\delta F_o$, $\delta 
G_o$, $\delta F_e$ and $\delta G_e$ are the leading order correction terms due 
to non zero electron mass. The observable $R_\eta$, now, gets modified as:
\small
\begin{align}
&\hspace{-3mm}{R}_\eta (x_n,y_n,q_n^2) 
=\frac{\overline{\Gamma}_o(x_n,y_n,q_n^2)}{\overline{\Gamma}_e(x_n,y_n,q_n^2)}
\nonumber\\
&\hspace{-3mm}\approx\bigg(\frac{{F}_o+(\frac{m_e}{m_\mu})^2\, \delta F_o}{{F}_e+(\frac{m_e}{m_\mu})^2\, \delta F_e}\bigg)\nonumber\\
&\times\bigg[1+\eta_\gamma\,
\bigg(\frac{{G}_o+(\frac{m_e}{m_\mu})^2\, \delta G_o}{{F}_o+(\frac{m_e}{m_\mu})^2\, \delta 
F_o}-\frac{{G}_e+(\frac{m_e}{m_\mu})^2\, \delta G_e}{{F}_e+(\frac{m_e}{m_\mu})^2\, \delta F_e}\bigg)\bigg]
\end{align}
\normalsize
Hence, the asymmetry, $A_\eta (x_p,y_p,q_p^2)$, in ${R}_\eta$ becomes, 
\begin{equation}
\begin{aligned}
\label{eq:expansion}
& A_\eta (x_n,y_n,q_n^2)=\dsp\Big(\frac{{R}_\eta}{{R}_{\rm SM}}-1\Big)\\[5pt]
&\approx\eta_\gamma\,
\bigg(\frac{{G}_o+(\frac{m_e}{m_\mu})^2\, \delta G_o}{{F}_o+(\frac{m_e}{m_\mu})^2\, \delta 
	F_o}-\frac{{G}_e+(\frac{m_e}{m_\mu})^2\, \delta G_e}{{F}_e+(\frac{m_e}{m_\mu})^2\, \delta F_e}\bigg)\\[5pt]
& \approx\eta_\gamma\,
\Big(\frac{{G}_o}{{F}_o}-\frac{{G}_e}{{F}_e}\Big)\\[5pt]
&\qquad+\eta_\gamma\,\Big(\frac{m_e}{m_\mu}\Big)^2\,\Big(\frac{G_e\,\delta
 F_e}{F_e^2}-\frac{G_o\,\delta F_o}{F_o^2}+\frac{\delta G_o}{F_o}-\frac{\delta 
G_e}{F_e}\Big)
\end{aligned}
\end{equation}
where,
\begin{equation*}
{R}_{\rm 
	SM}=\frac{\overline{\Gamma}_o}{\overline{\Gamma}_e}\bigg|_{\eta_\gamma=0}=\bigg(\frac{{F}_o+(\frac{m_e}{m_\mu})^2\, \delta F_o}{{F}_e+(\frac{m_e}{m_\mu})^2\, \delta F_e}\bigg).
\end{equation*} 
Note that the above expansion in ${\cal O}(m_e/m_\mu)$ fails in the region 
where collinear or soft photon divergences occurs.

\end{document}